\documentclass[namedreferences]{SolarPhysics}
\usepackage[optionalrh]{spr-sola-addons} 
\usepackage{graphicx}        
\usepackage{color}           
\usepackage{url}             
\usepackage{enumerate}

\newcommand{\aap}{    {\it Astron. Astrophys.}}

\newcommand{\apj}{    {\it Astrophys. J.}}

\newcommand{\pasj}{   {\it Pub. Astron. Soc. Japan}}

\newcommand{\solphys}{{\it Solar Phys.}}

\newcommand{\ssr}{    {\it Space Sci. Rev.}}

\begin{document}

\begin{article}

\begin{opening}

\title{Flare Hybrids}

%
\author{M.~\surname{Tomczak}\sep
        P.~\surname{Dubieniecki}
       }

%
\runningauthor{Tomczak and Dubieniecki} \runningtitle{Flare Hybrids}

%
  \institute{Astronomical Institute, University of Wroc{\l }aw,
  ul. Kopernika 11, 51-622 Wroc{\l }aw, Poland,
                     email: \url{tomczak@astro.uni.wroc.pl}\\
             }

\begin{abstract}
$\check{\rm{S}}$vestka (\textit{Solar Phys.} 1989, {\bf 121}, 399) on the basis of the {\sl Solar Maximum Mission} observations introduced a new class of flares, the so-called {\sl flare hybrids}. When they start, they look as typical compact flares (phase 1), but later on they look like flares with arcades of magnetic loops (phase 2). We summarize the features of flare hybrids in soft and hard X-rays as well as in extreme-ultraviolet; these allow us to distinguish them from other flares. Additional energy release or long plasma cooling timescales have been suggested as possible cause of phase 2. Estimations of frequency of flare hybrids have been given. Magnetic configurations supporting their origin have been presented. In our opinion, flare hybrids are quite frequent and a difference between lengths of two interacting systems of magnetic loops is a crucial parameter for recognizing their features.
\end{abstract}
%
\keywords{Flares, X-rays, Magnetic Fields}

\end{opening}

\section{Introduction}
     \label{intr}

There are not two identical flares, nevertheless it is useful to classify flares following some schemes. The most commonly accepted classification was introduced by \inlinecite{pal77} according to soft X-ray images obtained by the S-054 experiment on board {\em Skylab}. The authors proposed two separate classes of events, namely compact flares (class 1) and flares occurring in large and diffuse systems of loops (class 2). They found that the separation is supported by the different values of several physical parameters like height, volume, energy density, and characteristic times of rise, decay, and duration. They also perceived that flares of class 1 are located very low in active regions and, opposite to flares of class 2, do not appear to be associated with coronal mass ejections (CMEs) and prominence eruptions or activations.

The division into two classes opposed to each other by contradiction is called dichotomy, therefore we can shortly call the classification of \inlinecite{pal77} as the flare dichotomy. The flare dichotomy has been supported by several classifications, {\it e.g.} impulsive {\it vs.} long-duration flares, single-loop {\it vs.} arcade flares, confined {\it vs.} eruptive flares, or two-points {\it vs.} two-ribbons flares.

Beyond any doubt the division into two classes is very rough; therefore, it is possible that observed flares can share features of both class 1 and class 2. \inlinecite{sve89} introduced for them the term {\em flare hybrids}. How does a flare hybrid look like? Its evolution can be divided into two phases: during phase 1 it looks like a flare of class 1, and during phase 2 it looks like a flare of class 2. \inlinecite{sve89}\hspace*{-1.5mm}, recalling a private communication of Cornelius de Jager, discussed that a flare of class 1 may serve as a trigger of a flare of class 2. He also asked about the process which causes the magnetic field to open and thus start a flare of class 2.

Further observations made with many instruments at different wavelengths have derived the more complete picture of flare hybrids. In Section~\ref{obs} we will present characteristic features of flare hybrids in soft X-ray (SXR), hard X-ray (HXR), and extreme-ultraviolet (EUV) ranges, respectively. In Section~\ref{freq} some rules concerning a frequency of occurrence will be given. The magnetic configuration suggested for flare hybrids will be described in Section~\ref{conf}. In Section~\ref{conc} the most likely scenario for a flare hybrid will be proposed.

\section{Observations}
\label{obs}

\subsection{Soft X-rays}
\label{sxr}

\begin{figure}
\centerline{\includegraphics[width=1.1\textwidth]{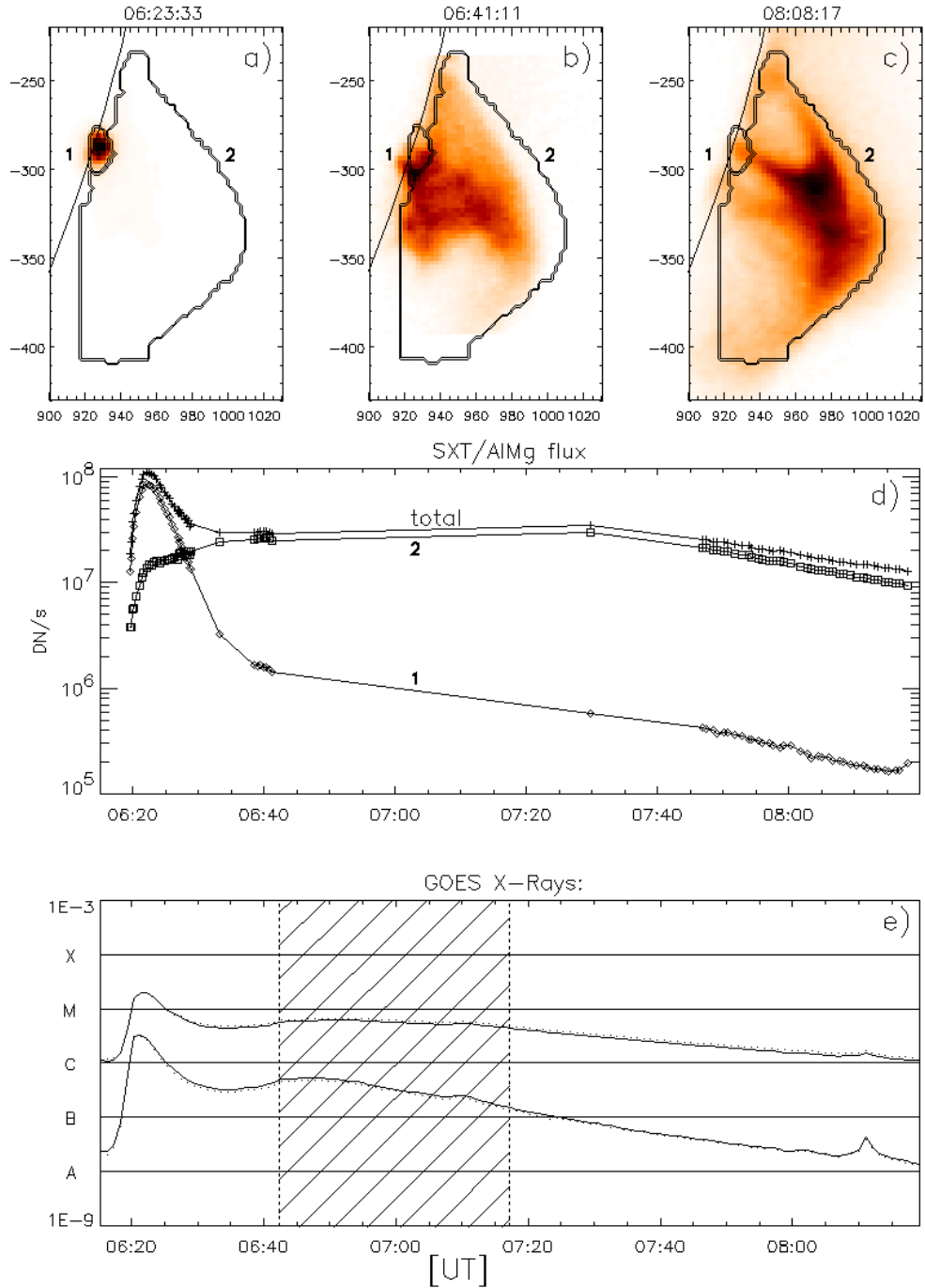}}
\hfill \caption{a)-c) The SXT/AlMg images of the flare hybrid SOL1992-11-05T06:22 (M2.0). The intensity scale is reversed. The solid line shows the solar limb, the double solid lines show the borders areas 1 and 2 shining in phase 1 and 2, respectively. d) The SXT/AlMg light curves for areas 1 (diamonds), 2 (boxes), and the total signal (crosses). e) The GOES light curves (upper curve -- 1\,--\,8 \AA\ range, lower curve -- 0.5\,--\,4 \AA\ range). The hatched areas show the {\sl Yohkoh} satellite nights. } \label{nov5s}
\end{figure}

In Figure~\ref{nov5s} we present an example of the flare hybrid, SOL1992-11-05T06:22 (M2.0), observed by the {\it Soft X-ray Telescope} (SXT: \opencite{tsu91}) on board the {\sl Yohkoh} satellite. The three SXR images made with the AlMg filter during phase 1 (Figure~\ref{nov5s}a), during phase 2 (Figure~\ref{nov5s}c), and in the intermediate time (Figure~\ref{nov5s}b) are given. As we can see, during phase 1 the SXR emission of the flare is dominated by a small ($h \approx 10^4$~km) system of bright loops. Later on, a higher magnetic arcade ($h \approx 5 \times 10^4$~km) is seen, which shines in SXRs during phase 2. In each image the borders of two areas, 1 and 2, within which the SXR signal appeared, are marked. Light curves for these areas, as well as the total signal from the full images, are presented in Figure~\ref{nov5s}d. Time gaps in the light curves are caused mainly by the satellite night. As we can see, light curves for areas 1 and 2 are different but together they compose a double-peak shape. The same double-peak light curve was recorded by the GOES satellites (Figure~\ref{nov5s}e), where the time interval of the satellite night is also marked.

\begin{figure}
\centerline{\includegraphics[width=1.1\textwidth]{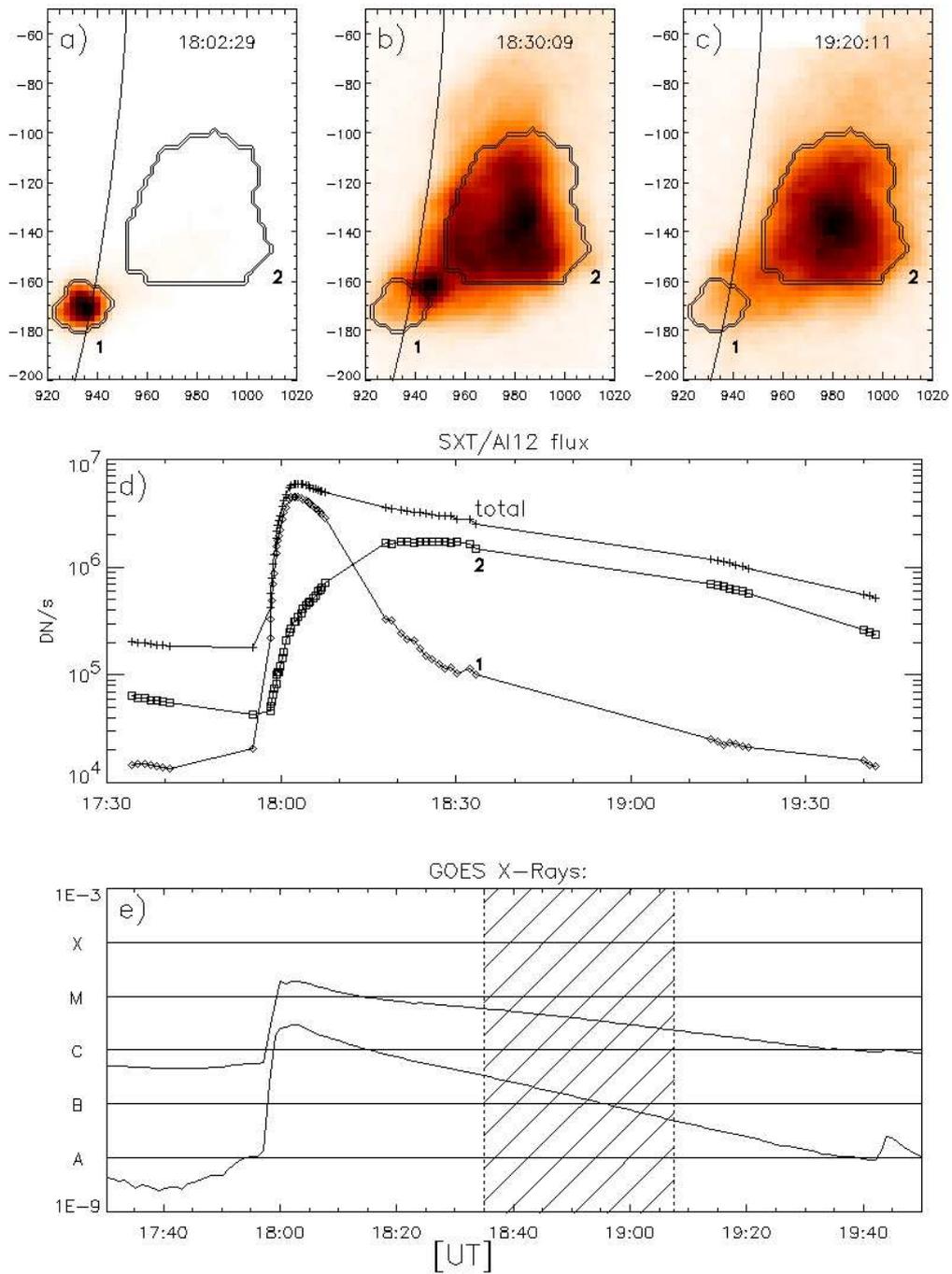}}
\hfill \caption{a)-c) The SXT/Al12 images of the flare hybrid SOL1992-09-09T18:03 (M1.9). d) The SXT/Al12 light curves for areas 1, 2, and the total signal. e) The GOES light curves. For more explanations, see caption to Figure~\ref{nov5s}.} \label{sep9s}
\end{figure}

However, a double-peak GOES light curve cannot be considered as a typical signature of a flare hybrid. Other example of a flare hybrid, SOL1992-09-09T18:03 (M1.9), observed by the SXT is given in Figure~\ref{sep9s}. The panels in the figure are organized in the same way as in Figure~\ref{nov5s}. The difference is the choice of another SXT filter, Al12. The evolution of the flare presented in Figure~\ref{sep9s} is very similar to that presented in Figure~\ref{nov5s}. During phase 1 the SXR emission of the flare is dominated by a smaller area 1 around a system of lower loops and during phase 2 an emission from a larger area 2 around a higher magnetic arcade dominates. Light curves for the areas 1 and 2 (Figure~\ref{sep9s}d) have their maxima shifted in time as in Figure~\ref{nov5s}d but this time they compose together an only one-peak light curve seen also in the GOES record (Figure~\ref{sep9s}e).

We investigated nine flare hybrids well-observed by {\sl Yohkoh} (Table~\ref{list}). In each case the evolution seen in SXT images looked very similar, namely during phase 1 the emission was concentrated in a system of rather small loops and during phase 2 the emission came from a larger arcade of loops. However, only the flare hybrid from Figure~\ref{nov5s} had a double-peak GOES light curve, while for the other events the GOES recorded one-peak light curves. Thus, we conclude that an intrinsic feature of a flare hybrid GOES light-curve is a rather strong asymmetry built by a short rise typical for flares of class 1 followed by a slow decay typical for flares of class 2. For events N$^0$ 1 to 6, 8, and 9 from Table~\ref{list} we found that the rise phase lasted 5 to 20 times shorter than the decay phase.

\begin{table}
 \caption{List of investigated flare hybrids}
 \label{list}
\begin{tabular}{cccccc}
\hline No. & Date & Max. time & GOES & Coordinates & NOAA AR \\
  & & [UT] & class & & \\
 \hline
1 & 30-Jan-92 & 17:15 & M1.6 & S12\,E84 & 7042 \\
2 & 8-Jul-92 & 09:50 & X1.2 & S11\,E46 & 7220 \\
3 & 11-Aug-92 & 22:28 & M1.4 & N16\,E90+ & 7260 \\
4 & 21-Aug-92 & 11:10 & M1.0 & N14\,W40 & 7260 \\
5 & 6-Sep-92 & 09:07 & M3.3 & S11\,W38 & 7270 \\
6 & 9-Sep-92 & 18:03 & M1.9 & S11\,W78 & 7270 \\
7 & 5-Nov-92 & 06:22 & M2.0 & S18\,W90+ & 7323 \\
8 & 9-Oct-93 & 08:11 & M1.1 & N11\,W78 & 7590 \\
9 & 22-Sep-97 & 14:16 & C4.7 & S28\,E43 & 8088 \\
 \hline
\end{tabular}
\end{table}

We used SXT data to calculate values of some parameters averaged over two systems of loops forming the investigated flare hybrids. For this aim we used the filter ratio method \cite{har92} employing the Be119 and Al12 image pairs as the first choice or the Al12 and Al.1 image pairs, when the Be119 images were not available. In this way we obtained a set of values characterizing the evolution of the temperature, $T$, and emission measure, $EM$. We estimated the total volume, $V$, of both components, hence values of the electron density number, $N_{\rm e}=\sqrt{EM/V}$, became available. Next, the total thermal energy, $E_{\rm th} = 3\,N_{\rm e}\,k_{\rm B}\,T\,V$ was calculated, where $k_{\rm B}$ is the Boltzmann constant. Finally, the heating rate {\it per} unit volume, $E_{\rm H} = (dE_{\rm th}/dt) + E_{\rm C} + E_{\rm R}$, was calculated, where $E_{\rm C}$ are the conductive losses and $E_{\rm R}$ are the radiative losses.

In Table~\ref{maxval} the maximum values for the parameters mentioned above are presented.  Moreover, the Be119 light-curve parameters are summarized in columns 2 to 4, where the time of maximum, maximum value of the signal, and its full width half maximum (FWHM) are given, respectively. We estimate that the typical values of the relative errors to be smaller than 2 to 3\% for the intensity, temperature, and emission measure. For the volume and the related parameters the relative errors are definitely larger, about 15 to 25\%. As we can see, in each case large amounts of energy were released in both components of the flare hybrids. Relations between parameters characterizing phase 1 and phase 2 of particular flare hybrids can be different, however, some trends are seen. Phase 1 always occurs in a smaller magnetic structure than phase 2 and lasts shorter. The smaller volume involved in phase 1 settles its larger electron number density, higher heating rate, and smaller total thermal energy than phase 2. Compare the mean values and their standard deviations given in the bottom rows of Table~\ref{maxval}.

\begin{table}
\caption{Maximum values of some parameters obtained for events from Table~1}
 \label{maxval}
\begin{tabular}{clrccrcrrc}
\hline No./ & Max. & $I_{max}$ & ${\Delta}t_{1/2}$ & $V$ & $\hspace*{-3mm}T$ & $EM$ & $N_{\rm e}$ & $E_{\rm th}$ & $E_{\rm H}$ \\
 phase & time & [10$^6$ & [min.] & [$10^{28}$ & [MK] & [$10^{49}$ & [$10^{10}$ & [$10^{30}$ & [ergs \\
  & [UT] & DN & & cm$^3$] & & cm$^{-3}$] & cm$^{-3}$] & ergs] & cm$^{-3}$ \\
  & & s$^{-1}$] & & & & & & & s$^{-1}$] \\
 \hline
1/1 & 17:14 & 1.2 & 15.5 & \hspace*{1mm}0.4 & 10.1 & 1.6 & 6.0 & 1.0 & 1.1 \\
1/2 & 17:20 & 1.1 & 25.2 & \hspace*{-0.5mm}10.4 & 10.2 & N/A & 1.1 & 4.6 & 0.3 \\
2/1 & 09:49.5 & 14.3 & \hspace*{1.5mm}5.3 & \hspace*{1mm}0.3 & 13.5 & \hspace*{-1.5mm}13.0 & 20.5 & 3.2 & \hspace*{-1.5mm}14.0 \\
2/2 & 10:04 & 2.0 & 19.0 & \hspace*{-0.5mm}10.9 & 16.4 & 3.2 & 1.8 & 10.8 & 1.2 \\
3/1 & 22:27.5 & 1.4 & \hspace*{1.5mm}3.5 & \hspace*{2mm}0.15 & 12.5 & 1.4 & 9.6 & 0.6 & 5.8 \\
3/2 & 22:34.3 & 1.4 & 17.7 & \hspace*{1mm}5.5 & 13.4 & 1.8 & 1.8 & 4.2 & 1.1 \\
4/1 & 11:06 & 0.5 & 11.5 & \hspace*{1mm}0.3 & 9.5 & 0.6 & 4.9 & 0.6 & 0.8 \\
4/2 & 11:14.5 & 1.0 & 32.3 & \hspace*{1mm}5.3 & 12.3 & 1.1 & 1.6 & 3.5 & 0.2 \\
5/1 & 09:06.5 & 3.8 & \hspace*{1.5mm}5.3 & \hspace*{1mm}0.3 & 13.3 & 4.5 & 13.0 & 1.5 & \hspace*{-1.5mm}10.2 \\
5/2 & 09:14 & 1.0 & 16.3 & \hspace*{1mm}3.6 & 13.2 & 1.4 & 2.2 & 3.5 & 0.9 \\
6/1 & 18:02.3 & 2.7 & \hspace*{1.5mm}8.2 & \hspace*{1mm}0.7 & 11.0 & 3.2 & 6.8 & 2.0 & 1.9 \\
6/2 & 18:21 & 2.6 & 56.5 & \hspace*{1mm}5.3 & 9.5 & 1.8 & 0.6 & 9.6 & \hspace*{2mm}0.05 \\
7/1 & 06:21.5 & 2.8 & \hspace*{1.5mm}4.5 & \hspace*{2mm}0.45 & 12.3 & 3.3 & 8.5 & 1.6 & 4.8 \\
7/2 & 06:41 & 1.4 & 91.0 & \hspace*{1mm}5.0 & 9.2 & N/A & 0.5 & 7.4 & \hspace*{2mm}0.07 \\
8/1 & 08:11.7 & 1.4 & \hspace*{1.5mm}5.8 & \hspace*{2mm}0.45 & 12.3 & 1.7 & 6.2 & 1.2 & 3.6 \\
8/2 & 08:19 & 0.6 & 19.8 & \hspace*{-0.5mm}12.8 & 11.6 & 0.8 & 0.8 & 4.1 & \hspace*{2mm}0.25 \\
9/1 & 14:16.3 & 0.5 & \hspace*{1.5mm}5.7 & \hspace*{2mm}0.15 & 10.8 & 0.6 & 6.4 & 0.4 & 2.8 \\
9/2 & 14:20 & 0.4 & 20.5 & \hspace*{1mm}3.8 & 11.5 & 0.5 & 1.2 & 2.0 & \hspace*{2mm}0.55 \\
 \hline
 \multicolumn{2}{l}{phase 1 (mean)} & 3.2 & 7.3 & \hspace*{2mm}0.36 & 11.7 & 3.3 & 9.1 & 1.3 & 5.0 \\
 \multicolumn{2}{l}{phase 1 (st. dev.)} & 4.3 & 3.9 & \hspace*{2mm}0.17 & 1.4 & 3.9 & 4.9 & 0.9 & 4.4 \\
 \multicolumn{2}{l}{phase 2 (mean)} & 1.3 & \hspace*{-3mm}33 & \hspace*{1mm}7.0 & 11.9 & 1.5 & 1.3 & 5.5 & 0.5 \\
  \multicolumn{2}{l}{phase 2 (st. dev.)} & 0.7 & \hspace*{-3mm}25 & \hspace*{1mm}3.4 & 2.3 & 0.9 & 0.6 & 3.0 & 0.4 \\
 \hline
\end{tabular}
\end{table}

Figures~\ref{nov5s}a-c and \ref{sep9s}a-c show that the areas labeled as 1 are situated near footpoints of the arcades 2. Moreover, the light curves in Figures~\ref{nov5s}d and \ref{sep9s}d show that the signal started to rise in the arcades 2 at the beginning of phase 1, which is not seen in Figures~\ref{nov5s}a and \ref{sep9s}a that are scaled to the brightest pixel. These facts strongly support the scenario -- in which the reported flares were not accidental coincidences of two events; the first one of class 1, and the second one of class 2, but rather a consequence of an interaction between the loop systems 1 and 2. We have observed similar behaviors for other investigated flare hybrids.

A fundamental question arises. Is phase 2 caused by additional energy release in the arcade or is it an effect of the long plasma cooling time within the larger structure? There is no doubt that magnetic reconnection between loop systems 1 and 2 can provide heating to both systems. The presence of reconnection is supported by intense HXR emission occurring during phase 1, when SXR emission strongly rises in both systems. There is also no doubt that conductive and radiative losses are higher in system 1 than in system 2 due to smaller sizes and higher electron number densities, respectively (see Table~\ref{maxval}). The higher energy losses should justify the shorter evolution timescales for system 1 in comparison with system 2, ${\tau}_1 \ll {\tau}_2$. However, the real evolution of flare loops is a complex interplay between heating and cooling processes.

\inlinecite{jak92} introduced the density-temperature ($N_{\rm e}-T$) diagram as a very useful diagnostic tool of heating process in a single flaring loop based on SXR observations. They showed that flare evolutionary paths during the decay phase on this diagram strongly depend on the duration of energy release. When the heating is switched off abruptly a cooling due to conductive and radiative losses quickly decreases the temperature causing a steep slope of the path ${\sim}2$. When the decay of the heating rate is rather slow, the cooling is much slower and the slope of the path is ${\sim}0.5$.

We investigated evolutionary paths of the analyzed flare hybrids on the $\sqrt{EM}-T$ diagram. As long as we built the paths with the data from the whole flare area, the paths seem to be more complicated than those obtained for simple hydrodynamic flare models \cite{jak92}. \inlinecite{syl93} interpreted the complicated evolutionary paths as a consequence of involving a set of distinct loops within the same flaring structure. The evolutionary paths on the $\sqrt{EM}-T$ diagram composed for both components of flare hybrids separately resemble a path modeled hydrodynamically for a single loop. Moreover, the decay slopes for the component 1 and the component 2 look usually very similar suggesting a slow decay of the heating rate. It should be stressed that the same phases in the evolutionary paths are shifted in time, for example, when we observe signatures of a prolonged energy release in the system 2 (a slope of the path ${\sim}0.5$), the evolutionary path for the system 1 is finished.

Unfortunately, phase 2 in the investigated flare hybrids lasted long enough to be interrupted by the satellite night. Further observations made during the next orbit in the late decay phase of flares suffered a lack of Be119 images which avoided a good temperature and emission measure diagnostics. For these reasons we cannot be sure that phase 2 of the investigated flare hybrids is caused by additional energy release in the system 2. Moreover, the available images do not allow us to identify a place of an additional reconnection.

\subsection{Hard X-rays}
\label{hxr}

\begin{figure}
\centerline{\includegraphics[width=0.9\textwidth]{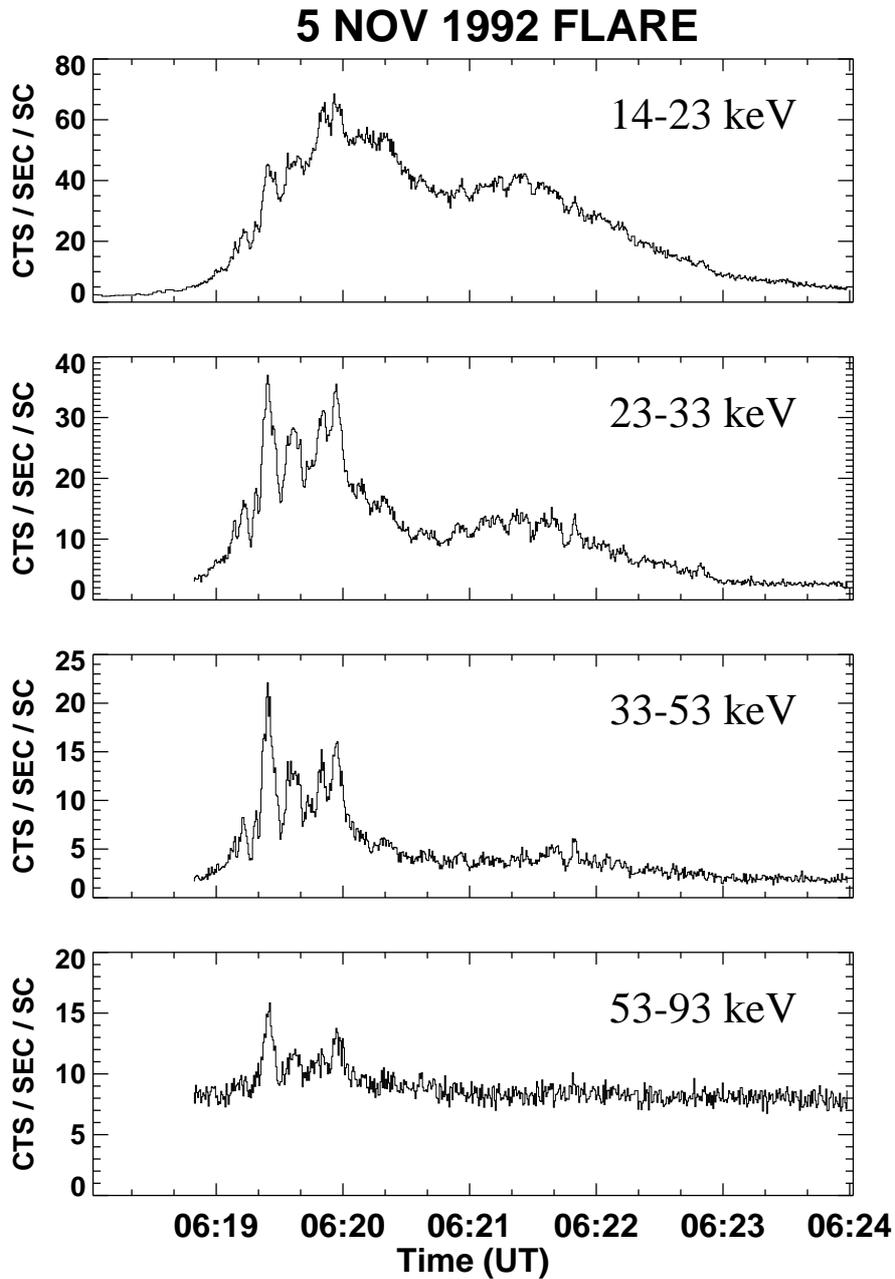}}
\hfill \caption{The {\sl Yohkoh} hard X-ray (HXR) light curves in four energy ranges. The number of counts are averaged {\it per} second and {\it per} subcollimator.} \label{nov5h}
\end{figure}

The HXR light curves in four energy channels for the flare in Figure~\ref{nov5s} (5 November 1992) are given in Figure~\ref{nov5h}. They were recorded by the {\it Hard X-ray Telescope} (HXT: \opencite{kos91}) on board the {\sl Yohkoh} satellite. In all the channels a sequence of almost equally spaced pulses, $P{\approx}13$~s, is seen. In lower energies (below 33 keV) this sequence lasts for about three minutes, between 06:19 and 06:22 UT, while in higher energies (above 33 keV) the pulses can be detected above the background only between 06:19 and 06:20 UT, due to lower count statistics. Similar pulsations, called quasi-periodic pulsations (QPP), are observed in many solar flares and it is commonly accepted that they are caused by MHD oscillations excited in flaring magnetic structures (\opencite{n+m09}, and references therein).

The pulses are modulated by an additional gradual component, which is well observed below 33 keV and absent above 53 keV. It consists of two broad enhancements lasting at least two minutes each, the maxima of which are separated in time by about 90~s (06:19:50 and 06:21:20 UT, respectively). The hardness ratio of a net signal above the background for two successive channels proves that the photon energy spectra of pulses are harder than the spectra of the gradual component.

Recently \inlinecite{t+s14} analyzed a similar solar flare of 2 October 2001 (SOL2001-10-02T17:31, C4.7), in which more energetic pulses with a period $P_1 = 26-31$\,s were modulated by three more gradual enhancements with a period $P_2 = 110$\,s. They found that these periods were excited simultaneously in a flare hybrid due to an interaction between a system of small loops and a high arcade of loops. They also proved that the shorter period was excited in the small loops and the longer period in the arcade.

The observations available for the flare of 5 November 1992 do not allow us to separate spatially both periods, but close similarity of the overall magnetic configuration with the flare of 2 October 2001 suggests an interaction between small loops and a high arcade exciting simultaneously by MHD oscillations in both magnetic structures. We would like to stress that in some articles reporting the presence of QPP with two distinctly different periods excited simultaneously in HXRs, an interaction between two magnetic structures of different sizes is always mentioned, {\it e.g.} \inlinecite{asa01}\hspace*{-1.5mm}, \inlinecite{nak06}\hspace*{-1mm}.

We investigated HXR light curves of all the flare hybrids from Table~\ref{list} looking for similarities with the flare of 5 November 1992, {\it i.e.} showing two distinctly different periods excited simultaneously. Unfortunately, some light curves were not complete due to the passage of the satellite through the South Atlantic Anomaly, therefore we incorporated the available light curves derived by the {\it Burst and Transient Source Experiment} (BATSE: \opencite{fis92}) on board the {\sl Compton Gamma Ray Observatory}. In summary, we confirm the presence of QPPs for seven flare hybrids, but only for three events two distinctly different periods were excited simultaneously. Apart from the flare of 5 November 1992, there were flares of 8 July 1992, 21 August 1992, and 9 September 1992. Four events from nine are for sure too few to claim that QPPs with two distinctly different periods excited simultaneously are intrinsic features of flare hybrids in HXRs. Nevertheless, this characteristic HXR pattern suggests an influence of the magnetic configuration of flare hybrids, in which two interacting magnetic structures have distinctly different sizes.

\subsection{Extreme-ultraviolet}
\label{euv}

\inlinecite{woo11}\hspace*{-1.5mm}, taking advantage of new instruments: the {\it Atmospheric Imaging Assembly} (AIA: \opencite{lem12}), and the {\it Extreme-ultraviolet Variability Experiment} (EVE: \opencite{woo12}) onboard the NASA {\sl Solar Dynamics Observatory} (SDO), introduced a new class of flares, called EUV late phase (ELP) flares. Their crucial feature is an additional peak of the coronal emission seen in the spectral line of Fe~{\sc xvi} 335\AA\ occurring half an hour to two hours after the GOES SXR peak. This line is an indicator of plasma with a temperature of about 3~MK. For other EUV spectral lines detecting the warm plasma, {\it e.g.} Fe~{\sc xviii} 94\AA\ ($\sim$6 MK) or Fe~{\sc xv} 284\AA\ ($\sim$2 MK), the late peaks also occur \cite{sun13}. \inlinecite{woo14} specified other features of ELP flares, namely: (1) no significant counterpart of the late peak in the hot plasma (the GOES SXR or Fe~{\sc xx}/Fe~{\sc xxiii} 133\AA\ ), (2) an eruption followed by a coronal dimming seen in the cooler plasma, {\it e.g.} Fe~{\sc ix} 171\AA\ , preceding the late peak, and (3) a different system of longer loops visible above the place where the peak of the hot plasma was emitted.

The features of ELP flares mentioned above strongly resemble those presented in Section~\ref{sxr} for flare hybrids observed in SXRs. The evolution of ELP flares consists of the two following phases: the first, when the emission is concentrated in a small system of loops (system 1), and the second, when an additional system of longer loops (system 2) occurs and its emission dominates. Moreover, an interaction between both systems undoubtedly exists. Therefore we consider the ELP flares simply as a new face of flare hybrids that we can investigate thanks to the new EUV instruments onboard the SDO. A broad wavelength range and high temporal, spatial, and spectral resolutions allow us to measure changes in many EUV spectral lines which are indicators of plasma in a wide range of temperatures. This gives us the opportunity to decide whether additional energy release or a long timescale of plasma cooling is responsible for the prolonged evolution.

In several case studies regarding the ELP flares \cite{hoc12,liu13,dai13,sun13} the authors investigated sequences of AIA light curves ordered with decreasing temperature of the filters from Fe~{\sc xx}/Fe~{\sc xxiii} 133\AA\ ($\sim$10 MK) to Fe~{\sc ix} 171\AA\ ($\sim$0.7 MK). The signal was formed by selected fragments of ELP flares. The authors agree that during phase 1 in the system 1 of the investigated flares the maximum is shifted in time, first occurring for the hot plasma, later on for the warm plasma, and finally for the cold plasma. This appearance is interpreted as a consequence of plasma cooling due to radiative and conductive losses. \inlinecite{liu13} reported a similar delay during phase 2 in the system 2. They obtained time scales definitely longer than for phase 1 due to less conductive (larger loops) and radiative (lower electron density number) losses. \inlinecite{sun13} reported similar results but only for the hot and warm plasma. The evolution of the cold plasma was more complex.

On the other hand, \inlinecite{dai13} obtained a more complicated picture, in which each light curve for the system 2 during phase 2 shows several maxima. They interpreted this observations as a proof of several episodes of energy release in this system. \inlinecite{hoc12} noticed a lack of the hot plasma in the system 2 during phases 1 and 2 and interpreted this fact as a proof of additional energy release in the system 2 causing only a modest increase of the temperature. The same interpretation can be used to explain double maxima visible in light curves of cold filters in the flare analyzed by \inlinecite{sun13}\hspace*{-1.5mm}.

Recently, \inlinecite{li14} have published additional arguments supporting a twofold explanation of phase 2 in ELP flares. They modeled the EUV emission from sets of loops having different lengths and different heating rates using the enthalpy based thermal evolution of loops (EBTEL) model \cite{kli08}. They found that two separate maxima in Fe~{\sc xvi} 335\AA\ can be modeled by simultaneous heating during phase 1 in two distinct loops of different lengths and by repeating the heating during phase 2 in the system 2. \inlinecite{li14} pointed out the importance of the AIA UV 1600\AA\ channel to distinguish between cooling and heating for the ELP flares. Some contribution to the emission in this channel comes from the C~{\sc iv} line formed in the upper chromosphere. Therefore the AIA UV 1600\AA\ channel is very sensitive to flare energy release. Indeed, the light curves presented by \inlinecite{li14} support the interpretation of additional energy release for the flares investigated by \inlinecite{dai13} and \inlinecite{sun13} as well as a lack of energy release in phase 2 of flares investigated by \inlinecite{liu13}\hspace*{-1.5mm}.

\section{How Common are Flare Hybrids?}
\label{freq}

We have investigated the occurrence of flare hybrids between September 1991 and April 1999. For this aim we analyzed SXR GOES light curves and qualified as flare hybrids (1) those showing a double maximum, under the condition that both maxima occurred in the same active region, or (2) those having a strongly asymmetric light curve, {\it i.e.} a fast rise and a slow decay. If possible, questionable events were verified on the basis of SXT images. In summary we identified 577 flare hybrids. Altogether in the investigated time interval 15178 flares occurred. It gives a 3.8\% contribution of flare hybrids.

\begin{figure}
\centerline{\includegraphics[width=1.1\textwidth]{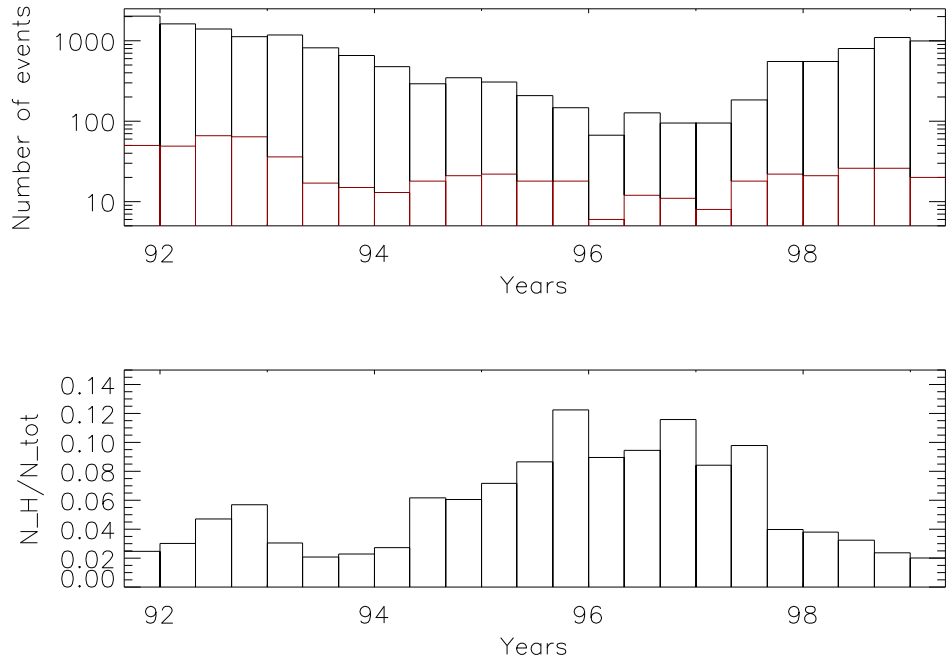}}
\hfill \caption{a) Number of flare hybrids (red bins) and all flares (black bins) that occurred between September 1991 and April 1999. The size of a bin is four months. b) Ratio of flare hybrids for each trimester bin.} \label{stat}
\end{figure}

Figure~\ref{stat} presents the number of flare hybrids for each trimester in the investigated time interval. For comparison the total number of all flares is also given. As we can see, the number of flare hybrids roughly mimics the solar cycle phase. However, the contribution of this class in the full population was the highest around the minimum of activity (1995-1997) $\approx$9-13\%, whereas during enhanced activity (1991, 1993, 1998-99) it lowered to $\approx$2-4\%.

Recently a more comprehensive statistical research concerning ELP flares has been done by \inlinecite{woo14}\hspace*{-1.5mm}. He investigated SXR GOES light curves from 1974 to 2013 looking for a dual-decay behavior, {\it i.e.} a steep slope followed by a moderate slope. He argued that the first slope represents cooling of shorter loops (system 1), whereas the second one the cooling of longer loops (system 2). \inlinecite{woo14} obtained that the contribution of flares showing the dual-decay is 7.9\%, in particular from 2010 to 2013 it was 10.5\%. He found that the contribution of ELP flares is the highest (20-30\%) around a solar-cycle minimum. He also found that the higher the flare class, the higher the contribution of the dual-decay flares. The very important work of \inlinecite{woo14} aimed to validate the results obtained with SDO data. He reported that 36\% and 57\% of flares showing the double-decay from 2010 to 2011 and from 2011 to 2013, respectively, does not show features allowing us to classify them as ELP flares.

In spite of fundamental differences between our and Wood's methodologies, there are also some similarities. The most striking is that in both studies the highest frequency occurs during low solar activity. The reason for this seems to be rather trivial. During low activity flares are not so frequent, thus their GOES light curves are not overlapped and every selecting criterion works perfectly. Keeping in mind that during the higher activity the probability of occurrence of active regions having complicated magnetic structure with loops of different lengths is even higher, the contribution of 9 to 13\% can be treated as a lower limit of flare hybrids. It is interesting that the values 20 to 30\% given by \inlinecite{woo14} for low activity phase, after adopting his 43\%-validation obtained for events from 2011 to 2013, decrease to 9 to 13\%.

\inlinecite{woo11} noticed that ELP flares show a tendency to occur within the same active regions. For example, their Table~2 informs that among from the 22 ELP flares, six events occurred in NOAA AR 11069 and another six in NOAA AR 11121. We did not complete the list of flare hybrids including the active region identification, but during the investigation of some active regions, within which a flare hybrid occurred, we found that the same magnetic configuration was flaring several times. For example, the flare hybrid described by \inlinecite{t+s14} was preceded by three other flare hybrids that occurred in the same active region NOAA AR 9628.

\section{Magnetic Configuration}\label{conf}

All the available observations of flare hybrids strongly suggest the existence of two sets of magnetically related loop systems. This means a multipolar magnetic configuration, in which magnetic reconnection plays a crucial role in energy release and in shaping a new configuration, that becomes more potential. In previous studies the following particular configurations have been proposed: a classical quadrupolar topology based on breakout reconnection \cite{hoc12}, an asymmetric quadrupolar topology with a sigmoidal core \cite{liu13}, a multi-step reconnection in a multipolar topology \cite{dai13}, and a fan-spine topology \cite{sun13}.

Usually we conclude about the reconnection indirectly by observing the emission of non-thermal electrons in HXRs and the emission of the multithermal plasma in SXRs and EUV. Sometimes eruptions can be treated as a signature of the reconnection and expanding loops can even initiate a following reconnection with a higher magnetic system \cite{su12,dai13}. Occasionally, it is possible to identify the loops that are the product of reconnection, {\it e.g.} see Figure~4 in \inlinecite{sun13}\hspace*{-1.5mm}. \inlinecite{tom13} reported a unique flare hybrid (SOL2001-10-02T17:31, C4.7), in which the reconnection occurred between a new emerging flux and an overlying coronal field. In SXR images recorded by the SXT the whole process is seen very well, starting from a fast expansion of emerging loops and their evident deformation in the vicinity of the reconnection site, followed by vigorous plasma motions inside the reconfiguring loops and the formation of a new system of loops.

The emerging flux model \cite{hey77}, in which subphotospheric magnetic fields emerge due to buoyancy within an already existing active region and meet overlying coronal magnetic fields easily explains the main characteristics of flare hybrids. Continuous emergence of a new flux under a stable magnetic environment can produce homologous flares occurring in the same location with similar morphologies. The examples mentioned in Section~\ref{freq} are likely the illustration of this process explaining the tendency of flare hybrids to occur in the same active region.

\section{Conclusions and Future Prospects}
\label{conc}

Our intention is to recall the forgotten term introduced by the late Professor Zden$\check{\rm{e}}$k $\check{\rm{S}}$vestka many years ago. His experience and intuition suggested the importance of flare hybrids that show a complex evolution in which their appearance changes completely. Events like these warn us against general conclusions formulated following a limited set of observations concerning, for example, an instant of time. The conclusion can be wrong even though in agreement with the available data. A closer insight needs sometimes the detailed study of the evolution of the active region in which the investigated event occurred.

For the last 25 years plenty of new data has been obtained by successive solar satellites. On this basis, one can easily recognize the following typical observational features of flare hybrids: (1) separate systems of loops seen in EUV and SXR images, (2) double-peaked or strongly asymmetrical light curves in these wavelengths, (3) multiperiodicity of pulses recorded in HXRs. Now it is possible to give comprehensive answers concerning the general questions asked by \inlinecite{sve89}\hspace*{-1.5mm}.

The conclusive condition for the occurrence of a flare hybrid seems to be the reconnection between two systems of magnetic loops, a system 1 and a system 2, having smaller and larger lengths, respectively. This means that the necessary condition for the occurrence is a multipolar magnetic configuration. The process can be initiated, for example, by a new magnetic flux (system 1) emerging within the already existing active region (system 2). The reconnection triggers energy release and the chromospheric evaporation which fills both systems with plasma initiating intense SXR and EUV emissions. The system 1 is quickly cooled, due to large radiative and conductive losses, completing phase 1. Further evolution of flare hybrids (phase 2) is connected with the evolution of the system 2. The prolonged SXR and EUV emissions in this system might be the effect of the long timescale of plasma cooling due to less radiative and conductive losses. However, in some events proofs of additional energy release in the system 2 are also given. A new reconnection site can be somehow connected with eruptions observed in phase 1. The amount of energy released in the system 2 during phase 2 establishes the maximum temperature of the plasma and in this way accessibility of observations in different SXR and EUV filters.

It is easier to recognize typical features of flare hybrids, when differences between lengths of the interacting systems of loops are larger. However, even very different systems do not always cause the clear features of these events. Therefore, the estimations of frequency of flare hybrids presented in Section~\ref{freq} should be treated as a lower limit of the actual value.

Flares are important for space weather due to the production of photons that are energetic enough to enhance quickly the ionization in Earth's upper atmosphere. Integration over wavelengths shows the energetic importance of the EUV flux. Therefore, the prolonged EUV emission makes the flare hybrids extremely geoeffective due to the prolonged impact on the Earth's atmosphere. For this reason, new methods introducing the early warning of occurrence of these events are welcome. Further complex investigation including X-ray and EUV observations should verify, for example, the usefulness of the multiperiodicity in HXRs for prediction of the EUV late maximum.

\begin{acks}
The {\sl Yohkoh} satellite is a project of the Institute of Space and Astronautical Science of Japan. We are very thankful to the referee for important comments which helped to improve this paper. We acknowledge financial support from the Polish National Science Centre grant 2011/03/B/ST9/00104.
\end{acks}

\end{article}

\end{document}